\documentclass[12pt]{article}
\usepackage{amsfonts,latexsym,fullpage,amsmath,latexsym,amssymb,epsf}

\date{March 28, 2002}

\newcommand{\ket}[1]{|#1\rangle}
\newcommand{\bra}[1]{\langle #1|}

\newcommand{\beq}{\begin{equation}}
\newcommand{\eeq}{\end{equation}}
\newcommand{\beqy}{\begin{eqnarray}}
\newcommand{\eeqy}{\end{eqnarray}}

\newtheorem{Definition}{Definition}
\newtheorem{Lemma}{Lemma}
\newtheorem{Theorem}{Theorem}

\newenvironment{Proof}{{\it Proof: \,}}{$\Box$ \vspace{0.3cm}}

\newenvironment{Definition*}{{\bf Definition}}{}

\def\C{{\mathbb{C}}}

\def\N{{\mathbb{N}}}

\newcommand{\B}{{\cal B}}
\newcommand{\cB}{{\cal B}}
\newcommand{\cH}{{\cal H}}

\begin{document}

\title{Cooling and Low Energy State Preparation
for $3$-local Hamiltonians are FQMA-complete}

\author{Dominik Janzing\thanks{e-mail: {\protect\tt
\{janzing,wocjan,eiss\_office\}@ira.uka.de}}, Pawel Wocjan, and
Thomas Beth \\ \small Institut f{\"u}r Algorithmen und Kognitive
Systeme, Universit{\"a}t Karlsruhe,\\[-1ex] \small Am Fasanengarten 5,
D-76\,131 Karlsruhe, Germany}

\date{February 4, 2003}

\date{March 28, 2003}

\maketitle

\abstract{We introduce the quantum complexity class FQMA. This class
describes the complexity of generating a quantum state that serves as
a witness for a given QMA problem. In a certain sense, FQMA is the
quantum analogue of FNP (function problems associated with NP). The
latter describes the complexity of finding a succinct proof for a NP
decision problem. Whereas all FNP problems can be reduced to NP, there
is no obvious reduction of FQMA to QMA since the solution of FQMA is a
quantum state and the solution of QMA the answer ``yes'' or ``no''.

We consider quantum state generators that get classical descriptions
of $3$-local Hamiltonians on $n$ qubits as input and prepare low
energy states for these systems as output. We show that such state
generators can be used to prepare witnesses for QMA problems. Hence
low energy state preparation is FQMA-complete. Our proofs are
extensions of the proofs by Kitaev et al.\ and Kempe and Regev for the
QMA-completeness of $k$-local Hamiltonian problems. We show that FQMA
can be solved by preparing thermal equilibrium states with an
appropriate temperature decreasing as the reciprocal of a polynomial
in $n$.}

\section{Introduction}
In analogy to classical computer science one may describe {\it
quantum} complexity theory as the field dealing with the complexity of
solving computational problems when a quantum computer is available.
However, in contrast to classical algorithms, the task of a ``quantum
algorithm'' is not necessarily to solve a {\it computational} problem.
It may, for instance, also be the task to prepare a certain
multi-particle quantum state which may be a resource for various
applications \cite{DowlingMilburn}. Since the preparation of generic
entangled states in multipartite systems is non-trivial, it makes
sense to consider the complexity of state preparation procedures
\cite{AharonovState}.

Remarkably, the question of complexity of state preparation
comes up straightforwardly when some {\it computational} problems
are generalized to the quantum domain.
This shows the following example:
Consider a function
\[
f: \{0,1\}^n \rightarrow \{0,1\}
\]
that is given by a boolean expression of polynomial size in $n$. The
task is to decide whether there exists an input $x$ such that
$f(x)=1$.  This problem is called SAT (Satisfiability). It is
NP-complete \cite{Papa}. The problem {\it to find} the input $x$
whenever it exists is called FSAT. In general one defines FNP as the
problem class that consists of the problems to find proofs to NP
decision problem.  It is easy to see that FSAT can be reduced to SAT
(with polynomial overhead) as follows: Define the functions $f^{0}$
and $f^{1}$ that are obtained if the first input bit is set to $0$ or
$1$, respectively and ask whether there exists an $x'$ of size $n-1$
such that $f^{0}$ has the output $1$ or that $f^{1}$ has (see
\cite{Papa}). This determines the first bit of the solution. In the
same way the other bits can be determined with a polynomial number of
calls of the decision problem\footnote{Note that there are
FNP-problems which are not {\it canonically} associated with
NP-problems.  They can be reduced to appropriate NP-problems
nevertheless \cite{Papa}.}.

The quantum analogue of this setting is as follows. Consider a
unitary quantum circuit $U$ of polynomial size (i.e. consisting of a
polynomial number of two-qubit gates) acting on
$(\C^2)^{\otimes(n+m)}$ where we have $n$ input and $m$ ancilla
qubits.  The ancilla register is initialized to the state
$\ket{0\ldots 0}$ in order to mimic irreversible mappings such as, for
instance, the classical logical functions AND, NAND, NOR, OR
\cite{BennettReversibel}.  According to \cite{KitaevShen} the quantum
analogue to SAT is as follows.  Given a quantum circuit $U$ that acts
on a tensor product of an input state $|\psi\rangle$ and the state
$|0\dots0\rangle$ of an initialized ancilla register. We obtain
\[
|\phi\rangle:=U (|\psi\rangle \otimes|0\dots 0\rangle)
\]
and measure the right most qubit. Then the quantum analogue to
``expression is not satisfiable'' is that for all states
$|\psi\rangle$ the measurement result is with high probability
``no''. The analogue to ``satisfiable'' is that there is a state
$|\psi\rangle$ such that the result is likely to be ``yes''.  This
class of problems is referred to as BQNP \cite{KitaevShen} or QMA
\cite{Watrous,KempeRegev} and each state $|\psi \rangle$ which leads
to ``yes'' with high probability is called ``witness'' of the
QMA-problem.

Note that the problem class QMA is only to {\it decide} whether there
exists a witness and not to prepare it if it exists. It should be
noted that we see no obvious reduction of the preparation problem to
the decision problem QMA.  Consider the case that there is an
entangled state $\ket{\psi}$ which is accepted by $U$ with high
probability. This does not imply that there is any state
$|\psi_1\rangle$ of the first qubit of the input register such that
$|\psi_1\rangle \otimes |\tilde{\psi}\rangle$ is accepted for an
appropriate state $|\tilde{\psi}\rangle$. Therefore, already the first
step of the reduction above fails. This may be a substantial
difference between quantum and classical: The class FNP given by the
set of problems to find a witness reduces to NP.  The problem to
prepare a ``witness'' of a QMA-problem shall be called FQMA here.
Note that it is not even clear whether there exists a polynomial
description of the witness state $|\psi\rangle$. Therefore FQMA is, in
the definition proposed here, not to {\it describe} the state that is
accepted, we demand rather to {\it prepare} it. In other words, the
answer of a QMA problem is classical whereas the answer of FQMA is
quantum.

In this paper we shall provide an example of an FQMA-complete problem
which has a rather intuitive physical meaning. It is the task to
prepare low-energy states in many-particle systems. This problem is
strongly related to the invention of efficient cooling mechanisms as
we will discuss later. The relation between efficient cooling of
many-particle systems with non-trivial interactions and solving hard
computational problems is well-known (see e.g.\
\cite{Paweladiabatic}). For Ising-type interactions one can show that
finding ground states requires to solve NP-complete problems
\cite{Pawelcompass}. However, these are examples where the solution is
only a classical spin configuration which is easy to describe. A
natural question is the following. Consider Hamiltonians for which it
is not clear whether a short classical description exists, is it a
hard quantum control problem to prepare low energy states? As usually
in complexity theory, we can only show that it is as hard as another
problem, namely solving FQMA in general.

\section{Introducing the complexity class FQMA}\label{SecFQMA}
In order to motivate our definition, we recall the definition of NP
and FNP problems. A language, i.e., subset of the set of strings is in
NP if there exists a polynomial-time decidable, polynomially balanced
relation $R_L$ such that there is a string $y$ with $R_L(x,y)$ if and
only if $x \in L$. The function problem associated with $L$, written
FL is the following computational problem:

Given $x$, return a string $y$ such that $R_L(x,y)$ provided that such
a string exists; if no such string exists, return ``no''. Less
formally it is the problem to give a proof that the answer is ``yes''
whenever it is true.

The complexity class QMA consists of the problems to decide whether a
given string is in a certain language in QMA. The set of QMA languages
is defined following \cite{KempeRegev}. First we introduce some
notations that will be used trough the paper. We denote the Hilbert
space of a qubit by $\B:=\C^2$. Let $x\in\{0,1\}^*$ be an arbitrary
binary string. We denote the length of $x$ by $|x|$. For any Hilbert
space $S(\cH)$ we denote the set of density matrices acting on $\cH$
by $S(\cH)$.

\begin{Definition}[QMA]\label{QMA}${}$\\
Fix $\epsilon=\epsilon(|x|)$ such that $2^{-\Omega(|x|)} \leq \epsilon
\leq 1/3$.  Then a language $L$ is in QMA if for every classical input
$x \in \{0,1\}^*$ one can efficiently generate (by classical
precomputation) a quantum circuit $U_x$ (``verifier'') consisting of
at most $p(|x|)$ elementary gates for an appropriate polynomial $p$
such that $U_x$ acts on the Hilbert space
\[
\cH:= \B^{\otimes {n_x}} \otimes \B^{\otimes m_x}\,,
\]
where $n_x$ and $m_x$ grow at most polynomially in $|x|$. The first
part is the input register and the second is the ancilla register.
Furthermore $U_x$ has the property that
\begin{enumerate}
\item For all $x \in L$
there exists a quantum state $\rho$ that is accepted by the
circuit with high probability, i.e.,
\[\exists \rho \in S(\B^{n_x})\,,\quad
tr(U_x\,(\rho\otimes\ket{0\ldots 0}\bra{0\ldots 0})\,U^\dagger_x\, P_1)
\geq 1-\epsilon\,,
\]
where $P_1$ is the
projection corresponding to the measurement ``Is the first qubit in
state $1$?''.
\item For all $x\not\in L$ all
quantum states are rejected with high probability, i.e.,
\[
\forall\rho \in S(\B^{n_x})\,,\quad
tr(U_x\,(\rho\otimes\ket{0\ldots 0}\bra{0\ldots 0})\,U_x^\dagger\, P_1)
\leq \epsilon\,.
\]
\end{enumerate}
\end{Definition}
Note that our ``proofs'' are mixed states in contrast to the
definitions in \cite{KitaevShen,KempeRegev}.  Due to linearity
arguments this modification does not change the language $L$.  Note
furthermore that it is always possible to construct a verifier for the
same language with $\epsilon'$ arbitrarily close to $0$. This
``amplification of probabilities'' is described in \cite{KitaevShen}
in full detail.  Since this amplification procedure has to be modified
for FQMA-problems we briefly sketch the idea. Use $k=poly(|x|)$ copies
of the circuit $U_x$. The decision whether the state $\rho'$ on
$(\B^{n_x} \otimes\B^{m_x})^{\otimes k}$ is accepted is based on a
kind of majority function considering all outcomes of measurements on
the first qubits of all the $k$ copies of the original register.  We
say that the total circuit accepts $\rho'$ if at least $k/2$
measurements return ``yes'' and rejects otherwise.  Actually, this
decision procedure does not fit strictly in the setting above since it
relies on $k$ measurements. This inconsistency can easily be removed
by realizing the majority vote by a ``small'' additional quantum
circuit. As shown in \cite{KitaevShen} there exists a joint state on
the $k$ copies of the original register that is accepted with high
probability if and only if there exists a state on one copy that is
likely to be accepted.

Now we define FQMA.
\begin{Definition}[FQMA]\label{FQMAdef}${}$\\
A channel with classical input $x$ and quantum output $\rho_x$ is in
FQMA if there is a language $L$ in QMA with a verifier $U_x$ and
$\epsilon$ as in Definition \ref{QMA} such that
\[
tr(U_x\, (\rho_x\otimes\ket{0\ldots 0}\bra{0\ldots 0})\,U^\dagger_x\,
P_1) \geq 1-\epsilon -\delta\,,
\]
whenever $x\in L$, where $\delta=1/r(|x|)$ for an arbitrary polynomial $r$.
For $x\not\in L$ the output is allowed to be arbitrary.
\end{Definition}

\section{Low energy state preparation\\ is FQMA-complete}
The determination of eigenvalues of non-trivial many-particle
interaction Hamiltonians is known as a computationally hard problem.
This is even the case if the Hamiltonians are restricted to $k$-local
ones. Here we call a Hamiltonian on $n$ qubits $k$-local if it is a
sum of operators that act only on $k$ qubits. Already for $2$-local
interactions one has NP-completeness \cite{Pawelcompass}. For
$3$-local terms it has recently been shown by Kempe and Regev
\cite{KempeRegev} (as extension of theorems by Kitaev et
al. \cite{KitaevShen}) that it is QMA-complete to decide whether the
Hamiltonian has an eigenvalue less than a given value.  More
explicitly, they constructed a $3$-local Hamiltonian associated with a
circuit $U$ (consisting of $L$ two-qubit gates) which has an
eigenvalue less than
\[
\frac{\epsilon}{L+1}
\]
if case 1 in Definition \ref{QMA} holds.
Conversely, all eigenvalues are at least
\[
\frac{c}{L^3}
\]
if case 2 holds, where $c$ is a constant. Moreover, Kitaev et al.\
\cite{KitaevShen} have shown that the problem to decide whether a
$k$-local Hamiltonian has an eigenvalue below $a$ or all eigenvalues
are above $b$ is in QMA whenever the gap $b-a$ is not stronger
decreasing than polynomially.

In the following we will show that the arguments in \cite{KempeRegev}
and \cite{KitaevShen} can be extended to prove that preparing low
energy states is FQMA-complete. Explicitly, we define this problem
class as follows.

\begin{Definition}[$k$-local Low Energy State Preparation]\label{Lowdef}${}$\\
Let $(H_n)$ be a sequence of $k$-local interaction Hamiltonians acting
on $\B^n$. The terms in the sum of each $H_n$ are assumed to be
positive and of norm not greater than $1$. Furthermore, they are
specified with $p(n)$ bits where $p$ is a polynomial. Let $(a_n)$,
$(b_n)$ and $(d_n)$ be sequences of numbers such that $b_n > d_n >
a_n$ and the gaps between $b_n$ and $d_n$ and between $d_n$ and $a_n$
decrease only polynomially in $n$. It is promised that either all
eigenvalues of $H_n$ are greater than $b_n$ or there exists an
eigenvalue smaller than $a_n$.

Then Low Energy State Preparation is the problem to prepare density
matrices $\rho_n$ on $n$ qubits (after $H_n$ has been specified by a
classical input) satisfying the low energy condition
\[
tr(\rho_n H_n) \leq d_n
\]
for all $n$ with $H_n$ having an eigenvalue less or equal than $a_n$.
\end{Definition}

\begin{Lemma}
The problem $k$-local Low Energy Preparation is in FQMA.
\end{Lemma}

\begin{Proof}
As Kitaev et al.\ \cite{KitaevShen} have shown there exists a language
in QMA such that $x\in L$ if there exist an eigenvalue smaller or
equal than $a_n$ and $x\not\in L$ if all eigenvalues are greater or
equal than $b_n$. We have to rephrase the verifier in order to show
that a state generator preparing witness states is also able to
prepare states with mean energy less or equal than $d_n$.

For fixed $n$ we drop the index $n$ and write $H$ as sum of its
$k$-qubit terms
\[
H=\sum_{j\leq r} H_j\,.
\]
Since $0<H_j<1$ for all terms, the operators $H_j, 1-H_j$ define a
positive operator valued measure. This POVM can be implemented with a
constant number of quantum gates, ancillas and one-qubit
measurements. Let the operator $H_j$ correspond to the result $0$ and
$1-H_j$ to the result $1$. Use a random generator that has one number
$1,\dots,r$ as outcome with equal probability. If the outcome is $j$
perform the measurement $H_j, 1-H_j$. The random generator can be
implemented by an appropriate quantum circuit and the measurements can
be replaced by controlled bit-flips on an additional ancilla which
shows the measurement result. This defines a quantum circuit $U$ and
the latter ancilla qubit is the qubit that is measured. Thus we have
the setting of Definition \ref{FQMAdef}. Then the probability that a
state $\rho$ is rejected is given by
\[
tr(\frac{1}{r} H_j)
\]
Assume there exists a state with energy at most $a_n$. Then it is
accepted by the circuit with probability $1-a_n/r$. A channel which
is a ``universal FQMA solver'' is able to prepare a state which is
accepted with probability at least $1-d_n/r$, i.e., with mean energy
at most $d_n$.
This completes the proof.
\end{Proof}

As we will prove in the following, we have also the other direction:
\begin{Theorem}[FQMA-completeness]\label{Main}${}$\\
The Problem $3$-local Low Energy Preparation is FQMA-complete, i.e.,
each quantum state generator that solves the $3$-local Low Energy
Preparation Problem can be used to generate witnesses for QMA problems
with arbitrary precision. Whenever there exists a state that is
accepted by a given circuit with probability $1-\epsilon$ the ``Low
Energy State Generator'' can be used to prepare a state that is
accepted with probability $1-\epsilon-\delta$ where $\delta$ can be
made arbitrarily small. The complexity of the procedure is
$O(1/\delta^4)$.
\end{Theorem}

Given a quantum circuit $U$ (note that we shall drop the subscript $x$
in the following), the task is to construct a Hamiltonian such that
its low energy states can efficiently be transformed into a state that
is accepted by $U$ with high probability.

The constructions of Kitaev et al. \cite{KitaevShen} and Kempe and
Regev \cite{KempeRegev} show a correspondence between states that are
accepted by a given circuit and low energy states of the corresponding
Hamiltonian.  One may therefore expect that our Theorem is already
given by a straightforward reinterpretation of their results.
However, the main problem is that such a straightforward approach
allows only to prepare states that are accepted with considerably
lower probability than $1-\epsilon$. The QMA problem is to decide
whether there exists a state that is accepted with probability at
least $1-\epsilon$ or whether all states are accepted with probability
at most $\epsilon$. The gap between $\epsilon$ and $1-\epsilon$ is
essential for the proofs in \cite{KitaevShen,KempeRegev}. We cannot
make use of such a gap. Whenever there exist states that are accepted
with probability $1-\epsilon$ and others with probability $\epsilon$
there are, by linearity, always states that are accepted with
probability $p$ for every value between $1-\epsilon$ and
$\epsilon$. In the straightforward extension of \cite{KitaevShen} and
\cite{KempeRegev}, it is not clear how to avoid that some of these
states are obtained. Therefore we demand only to construct states
which are accepted with probability $1-\epsilon-\delta$. As we will
see below, the allowed inaccuracy $\delta$ is essential for our
construction.

The main idea is to construct a circuit $\tilde{U}$ which is a
modification of the amplification explained in Section \ref{SecFQMA}.
For the ``meta-circuit'' $\tilde{U}$ exist states that are accepted
with probability $1-\tilde{\epsilon}$ where $\tilde{\epsilon}$ is
considerably smaller than $\epsilon$. Now we can use the construction
of \cite{KempeRegev} and obtain a Hamiltonian. We show that its low
energy states can be transformed simply to states are accepted by
$\tilde{U}$ with high probability. Although the probability of
acceptance may in general be considerably smaller than the possible
value $1-\tilde{\epsilon}$ we can use them to obtain states which are
accepted by the {\em original} circuit $U$ with probability almost
$1-\epsilon$.

To motivate our construction of $\tilde{U}$ one should note that
Kitaev's probability amplification does not provide states that are
accepted by the original circuit $U$ with high probability. Assume,
for instance, that we use $k=3l$ copies of $U$ for large $l\in \N$.
Let the input of $\tilde{U}$ be the state
\begin{equation}\label{joint}
\rho^{\otimes 2l} \otimes \gamma^{\otimes l}\,,
\end{equation}
where $\rho$ is a state that is always accepted by $U$ and $\gamma$ is
a state that is never accepted. The joint state (\ref{joint}) is very
likely to be accepted by the majority vote. A straightforward method
to obtain an input state for the original circuit would be to choose
randomly the restriction of the joint state to one of the $k$ input
registers. We would get the state
\[
\frac{2}{3} \rho + \frac{1}{3} \gamma\,.
\]
This state is rejected with probability $1/3$ independent of $k$.

Therefore we modify the ``majority vote'' in such a way that we accept
the joint state $\tilde{\rho}$ only if more than $k(1-\epsilon -
1/\sqrt[4]{k})$ of the original circuits $U$ accept their
corresponding parts of $\tilde{\rho}$. By choosing randomly one of the
$k$ reduced states one obtains a mixed state that is accepted by $U$
with conditional probability at least
\[
1-\epsilon -1/\sqrt[4]{k}\,,
\]
given the event that the meta circuit has accepted $\tilde{\rho}$.

The following lemma shows quantitatively that there exists a state
that is accepted by the meta circuit with high probability:

\begin{Lemma}
Given the modified majority vote which accepts input states of
$\tilde{U}_x$ whenever more than $k(1-\epsilon - 1/\sqrt[4]{k})$
copies of $U_x$ have accepted.  Let $\rho$ be a state that is accepted
by $U_x$ with probability at least $1-\epsilon$.  Then a $k$-fold copy
of $\rho$ is rejected with probability
\[
O( 2^{- \sqrt{k}/ \ln 2})\,,
\]
i.e. the probability can be amplified {\em sub-exponentially}.
\end{Lemma}

\begin{Proof}
Assume the worst case that $U$ accepts $\rho$ exactly with probability
$1-\epsilon$. In order to have a lower bound on the probability that
$\rho^{\otimes k}$ is accepted we have to estimate the tail of a
binomial distribution according to a Bernoulli experiment with
probability $\epsilon$:
\[
P(\rho^{\otimes k} \hbox{ rejected }) = \sum_{j \leq l} { k \choose j
} \epsilon^j (1-\epsilon)^{k-j}=: \sum_{j \leq l} b_k(j) \,,
\]
where
\[
l:=\lceil k(1-\epsilon -1/\sqrt[4]{k}) \rceil\,.
\]
For $l\geq k/2$ we have certainly $\sum_{j\leq l} b_k(j) \leq l\,
b_k(l)$.  The binomial coefficient $b_k(l)$ can be estimated using
relative entropies \cite{Blake}:
\[
b_k(l) = O( 2^{-k D(l/k)})
\]
with
\[
D(l/k):=(l/k) \log_2 \epsilon -(l/k) \log_2 (l/k) + (1-l/k)\log_2
(1-\epsilon) -(1-l/k) \log_2( 1-l/k)
\]
Note that the term $D(l/k)$ is the Kullback-Leibler relative entropy
\cite{Cover} which measures the distance between the probability
measure on two points given by $(\epsilon, 1-\epsilon)$ and the
(formal) probability measure defined by the observed relative
frequencies $(l/k,1-l/k)$.

Hence $\rho^{\otimes k}$ is rejected with probability
\[
O(2^{-k D(1-\epsilon -1/\sqrt[4]{k})})\,.
\]
We use Taylor expansion around the point $1-\epsilon$ up to the first
order with an appropriate remainder term and obtain
\[
D(1-\epsilon -1/\sqrt[4]{k})=D(1-\epsilon)+ D'(1-\epsilon)
1/\sqrt[4]{k} +D''(\Theta) \frac{1}{2(\sqrt[4]{k})^2} \,,
\]
for an appropriate value $\Theta$ in the interval $(1-\epsilon,
1-\epsilon + 1/\sqrt[4]{k})$. Relative entropy is always strictly
positive for two different probability measures. Therefore the
function $D$ has its unique minimum at $1-\epsilon$ and the first and
second term in the expansion vanish. The term $D''(\Theta)$ is at
least $2/\ln 2$. We have
\[
D(1-\epsilon -1/\sqrt[4]{k})\geq \frac{1}{\ln 2 \sqrt{k}}\,.
\]
We conclude
\[
P(\rho^{\otimes k} \hbox{ rejected }) = O( 2^{- \sqrt{k}/ \ln 2})\,.
\]
\end{Proof}

Now we are able to prove Theorem \ref{Main}. The first part of the
proof rephrases the constructions and notations of \cite{KitaevShen}
and \cite{KempeRegev}.
\begin{Proof}
Let $U=U_L U_{L-1}\dots U_1$ consist of $L$ two-qubit gates.  The
circuit $U$ acts on $\cB^l=\cB^n \otimes \cB^m$ qubits, where $n$ and
$m$ are the size of the input and the ancilla register, respectively.
Furthermore we need an additional register called ``clock''. It
consists of $L$ qubits. The total Hamiltonian consists of four parts:
\[
H:=H_{in} +H_{out} + H_{prop} + H_{clock}\,.
\]
The low energy states of $H$ represent in some sense the whole history
of the quantum circuit $U$. The correlations between the clock register
and the Hilbert space the circuit acts on contain the information at
which time step which gate has been applied.  This can be achieved as
follows.
\[
H_{clock}:= L^{12} \sum_{1\le i < j\le L}
\ket{01}_{ij}\bra{01}_{ij}\,.
\]
The subscripts $i,j$ indicate the considered qubits. $H_{clock}$ acts
only on the $L$ clock qubits. It penalizes all states in the clock
register that are not of the form
\[
\ket{\underbrace{11\ldots 1}_{t}\underbrace{0\ldots 0}_{L-t}}\,,
\]
called ``unary representation'' of the numbers $1,\dots,L$.
States of this form are denoted by $|\hat{t}\rangle$.
They are the allowed states of the clock.

Consider $H_{in}$. It is defined as follows.
\[
H_{in}:=\sum_{i=m+1}^N |1\rangle_i \langle 1|_i
\otimes |0\rangle_1 \langle 0|_1
\]
The first component of the tensor product acts on the ancilla register
and the second on the first qubit of the clock. It penalizes all
states where the ancilla register is not initialized whenever the
clock is in its starting position.

Consider $H_{prop}$. It ensures that the correlations between the
clock and the remaining registers are according to the history of the
quantum circuit.
\[
H_{prop}:=\sum_{t=1}^L H_{prop,t}
\]
with
\[
H_{prop,t}:= \frac{1}{2} (
1 \otimes \ket{10}_{t,t+1} \bra{10}_{t,t+1} +
1 \otimes \ket{10}_{t-1,t} \bra{10}_{t-1,t} -
U_t \otimes \ket{1}_{t} \bra{0}_{t} +
U_t^\dagger \ket{0}_{t} \bra{1}_{t} )
\]
for $2\leq t \leq L-1$ and
\begin{eqnarray*}
H_{prop,1}:= \frac{1}{2} (
1 \otimes \ket{10}_{1,2} \bra{10}_{1,2} +
1 \otimes \ket{0}_{1} \bra{0}_{1} -
U_1 \otimes \ket{1}_1 \bra{0}_1 +
U_1^\dagger \ket{0}_1 \bra{1}_1) \\
H_{prop,L}:= \frac{1}{2} (
1 \otimes \ket{1}_{L} \bra{1}_{L} +
1 \otimes \ket{10}_{L-1,L} \bra{10}_{L-1,L} -
U_L \otimes \ket{1}_{L} \bra{0}_{L} +
U_L^\dagger \ket{0}_{L} \bra{1}_{L}) \,.
\end{eqnarray*}

Consider $H_{out}$. It penalizes all states where the output is not
in the state $|0\rangle$ whenever the clock is in its end position.
\[
H_{out}:=|0\rangle_1 \langle 0|_1 \otimes |1\rangle_L \langle 1|_L\,.
\]

We also need unitary transformation $W$ defined in \cite{KitaevShen}
\[
W:= \sum_{t=0}^L U_t U_{t-1} \cdots
U_1 \otimes |\hat{t}\rangle \langle \hat{t}|\,.
\]
for transforming the low energy states. This transformation can
obviously be implemented efficiently since it can be written as
product of transformations
\[
W_t:= U_t\otimes |1\rangle_t \langle 1|_t\,,
\]
where each $W_t$ is a $3$-qubit gate.

Having introduced the necessary notation we come to the second part of
the proof. Let $\rho$ be a state such that $tr(\rho H) \leq a$ and
$\sigma$ be the restriction of the state $W^\dagger \rho W$ to the
input register. We show that $\sigma$ is accepted by the circuit $U$
with probability at least
\[
1 - a^{1/4} \,poly(L)\,.
\]
Here and in the rest of the proof we write $poly(L)$ for every
expression $f(L)$ which is smaller than an appropriate polynomial in
the size $L$ of the circuit independent of its specific structure and
independent of $a$.

Without loss of generality we may assume $\rho$ to be pure due to
linearity of mean energy.
Let $W^\dagger \rho W$ be the state $|\phi \rangle \langle \phi|$.
We decompose $|\phi\rangle$ as follows
\[
|\phi \rangle = d_0 |\alpha_0 \rangle \otimes |\psi_0 \rangle
+ \sum_{j\geq 1} d_j |\alpha_j\rangle \otimes |\psi_j \rangle\,,
\]
where $|\psi_j\rangle$ is the $j$-th eigenvector of
$W^\dagger(H_{prop} +H_{clock})W$ (starting with $0$). The only
eigenvector corresponding to the eigenvalue $0$ is
\[
|\psi_0 \rangle:= \frac{1}{\sqrt{L+1}} \sum_t |\hat{t}\rangle\,.
\]
The other eigenvalues are bounded from below by the non-zero
eigenvalues of $H_{prop}$. They are given by $\lambda_k=1-\cos q_k$
with $q_k:=\pi k/ (L+1)$. Asymptotically, the eigenvalues are larger
than $c/L^2$ for an appropriate positive constant $c$. We have
\[
a \geq 
tr( W^\dagger \rho W W^\dagger H W) = 
\langle \phi | W^\dagger HW | \phi \rangle \geq 
\langle \phi | W^\dagger H_{prop} W| \phi \rangle \geq 
\sum_{j\geq 1} |d_j|^2 q_j \,.
\]
We conclude
\[
\sum_{j \geq 1} |d_j|^2 \leq a \,poly(L) \,.
\]
Now we neglect the terms for $j>0$. The remaining state is
\[
|\gamma \rangle :=|\alpha_0 \rangle \otimes |\psi_0\rangle \,.
\]
The norm distance between $|\phi\rangle$ and $|\gamma\rangle$ can now
be estimated by elementary geometry. Let $\ket{x}$ and $\ket{y}$ be two
vectors such that $\ket{z}=\ket{x}+\ket{y}$ is a unit vector. Then we always
have 
\[
\| \ket{z} - \frac{1}{\| |x\rangle \|} |x\rangle \| \leq \sqrt{2}
\||y\rangle\|\,,
\]
since the left hand side is the length of a hypotenuse of an rectangular
triangle with legs of length $\| |y\rangle \|$ and
$1-\| |x\rangle \| < \| |y\rangle \|$.

Hence the norm distance between $|\gamma\rangle$ and the true state
vector $|\phi\rangle$ is less than $\sqrt{2 \sum_{j\geq 1} |d_j|^2} =
\sqrt{a} \,poly(L)$

We decompose $|\gamma \rangle$ as
\[
|\alpha_0 \rangle \otimes |\psi_0\rangle=
c_0 |\eta_0 \rangle \otimes | 0\dots 0\rangle \otimes |\psi_0\rangle +
\sum_{1 \leq b \leq 2^{m}} c_b |\eta_b\rangle \otimes |b\rangle \otimes
|\psi_0\rangle \,,
\]
where $|b\rangle$ denote the basis states of the ancilla register and
$|\eta_b\rangle$ are arbitrary state vectors in the input register.

In the following we will use the following elementary argument several
times: If the norm distance between two unit vectors $|x\rangle$ and
$|y\rangle$ is $\Delta<1$ one has
\[
\langle x| A | x\rangle - \langle y| A| y\rangle| \leq \|A\|3\Delta\,,
\]
where $A$ is an arbitrary matrix with operator norm $\|A\|$.  Using
this estimation we have
\[
a+ \sqrt{a} \, poly(L) \|H\| \geq
\langle \gamma | W^\dagger H W |\gamma\rangle\geq
\langle \gamma | W^\dagger H_{in} W |\gamma \rangle
=\frac{1}{L+1}\sum_{1\leq b \leq 2^m} |c_b|^2 \,.
\]
Since $\|H\| \leq poly(L)$
we find
\[
\sum_{1\leq b\leq 2^m} |c_b|^2 \leq \sqrt{a}\, poly(L)\,.
\]

We neglect the part of the state $|\gamma\rangle$ with $c_b$ for
$b\geq 1$. The remaining state has the form
\[
|\chi \rangle :=|\eta_0 \rangle \otimes |0\dots 0\rangle \otimes
|\psi_0\rangle\,.
\]
Its norm distance from $|\gamma\rangle$ is $a^{1/4} poly(L)$.
Its mean energy is at least
\[
\langle \chi | W^\dagger H_{out} W|
\chi \rangle
=\frac{P( |\chi\rangle \hbox{ rejected })}{L+1}\,.
\]
The norm distance between $|\phi\rangle$ and $|\chi\rangle$
is $a^{1/4} \,poly(L)$.
By the geometry argument above
and $\|H_{out}\|=1$ we have
\[
a\geq \langle \phi | W^\dagger H_{out} W |\phi \rangle \geq
\frac{P( |\chi\rangle \hbox{ rejected })}{L+1} - a^{1/4} \,poly(L)
\]
This implies
\[
P( |\chi\rangle \hbox{ rejected }) \leq a^{1/4} \, poly(L)\,.
\]
This shows that for each state $|\phi\rangle$ with energy $a$ the
corresponding state $|\chi\rangle$ is rejected with probability
$a^{1/4} poly(L)$. Since the norm distance between $|\chi\rangle $ and
$|\phi\rangle$ is $ a^{1/4} poly(L)$ we conclude that each state with
energy smaller than $a$ is rejected with probability at most
\begin{equation}\label{Schranke}
\epsilon':= a^{1/4} \, poly(L) \,.
\end{equation}
The Hamiltonian $H$ corresponding to the meta circuit $\tilde{U}$ as
explained in the first part of the proof has an eigenvalue of the
order
\[
O(2^{-\sqrt{k}/\ln 2})\,.
\]
Due to \cite{KitaevShen} and \cite{KempeRegev} this is guaranteed by
the fact that $\tilde{U}$ has an input that is rejected with this
small probability. Hence we can use the Low Energy State Generator for
preparing a state $\rho$ with energy $d\le 1/s(L)$ for any polynomial
$s$.

Let $\hat{\rho}$ be the restriction of $\rho$ to a randomly chosen
sub-register. The conditional probability that it is accepted by $U$
given the event that $\rho$ has been accepted by the meta circuit
$\tilde{U}$ is at least
\[
1 - \epsilon - \frac{1}{\sqrt[4]{k}}
\]
as already explained above. The meta circuit accepts with probability
$1 - d^{1/4} \, poly(L)$. The unconditional probability that
$\hat{\rho}$ is accepted is at least
\[
(1- \epsilon - \frac{1}{\sqrt[4]{k}})(1-d^{1/4}\,poly(L)) >
1-\epsilon -\frac{1}{\sqrt[4]{k}} - d^{1/4}\, poly(L)
\]
Now we have to show that
\[
\frac{1}{\sqrt[4]{k}} + d^{1/4}\, poly(L)
\]
can be made smaller than $\delta$. Choose $k$ such that
$1/\sqrt[4]{k}$ is smaller than $\delta/2$, i.e., $k> 16/\delta^4$.

As already noted, it is possible to choose $d$ such that $d^{1/4}\,
poly(L) < \delta/2$. To see this, recall that $L$ is here not the
length of the original circuit but of the meta circuit
$\tilde{U}$. However, it increases only polynomially in $k$ and the
least possible value of $a$ is $O(2^{-\sqrt{k}/\ln 2})$ decreasing
faster than the reciprocal of every polynomial. Therefore $d$ can be
chosen as required.
\end{Proof}

\section{Which temperature is required \\
for solving FQMA and QMA?}  

Obviously low energy state preparation can be achieved by cooling.
But clearly the invention of efficient cooling mechanism is a highly
non-trivial problem for theoreticians and experimentalists. In the
context of quantum computation algorithms have been suggested to
prepare low energy states on a system with universal quantum
computation capabilities \cite{AharonovState,TerhalDiVin}. At first
sight there seems to be a fundamental difference between the task to
cool a system with given Hamiltonian and to prepare a thermal
equilibrium state according to a Hamiltonian that is specified by
classical information only. In the latter case the Hamiltonian is only
{\it virtual}, i.e., not physically present. However, from the point
of view of complexity theory this does not make any difference: It
seems that every cooling mechanism can be translated into an algorithm
that prepares low temperature states on a quantum computer: Given the
postulate that every physical process can efficiently be simulated on
a quantum computer one can clearly simulate the time evolution
according to the Hamiltonian of the system that is to be cooled and
the interaction between this system and its environment (the entropy
sink). Whether each algorithm can be translated into a cooling
mechanism depends on the control operations which are available.
Consider for instance an $n+k$-qubit system with pair-interactions
forming a connected graph. Assume furthermore that all one-qubit
transformations can be implemented on a time scale that is smaller
than the time evolution according to the interaction. Then
decoupling-techniques can be used to ``switch off'' unwanted
interactions or transform them to different interaction types in such
a way that universal quantum computation is possible
(e.g. \cite{DNB,GraphPawel,BCL}). Hence one can implement every
quantum algorithm which transports entropy from the $n$ qubits to the
$k$ ancilla qubits.

The efficiency of algorithms preparing equilibrium states seems to
depend strongly on the demanded temperature
\cite{TerhalDiVin,AharonovState} and also the required thermodynamical
resources increase strongly with temperature tending to the absolute
zero \cite{JWZ00}.

We consider the question which temperature is needed to solve FQMA.
Let us first consider Low Energy State Preparation (see Definition
\ref{Lowdef}). The gap between the energy value $d_n$ to be achieved
and the ground state energy $g_n$ is essential for the required
temperature. To see this define a cutoff energy value by
\[
c_n:= a_n + \frac{1}{2} (d_n - a_n)\,.
\]
In the following we shall drop the index $n$. Consider the Gibbs
equilibrium state $\rho$ with absolute temperature $T$. Let
$(|\psi_j\rangle)_j$ be a basis of eigenstates of $H_n$ with
corresponding eigenvalues $(E_j)_j$. Then we have
\[
\rho:= \sum_j p_j |\phi_j \rangle \langle \phi_j|
\]
with
\[
p_j := \frac{e^{-E_j / (KT)}}{\sum_j e^{ -E_j / (KT)}}\,,
\]
where $K$ is the Boltzmann constant. We estimate the probability
$p_j$ for all $j$ with $E_j > c$. We have
\[
p_j < \frac{e^{-E_j/(KT)}}{ e^{-g/(KT)} + e^{-E_j/(KT)}}
\leq \frac{1}{e^{(c-a)/(KT)}}
\]
We have at most $2^n-1<2^n$ states above $c$. Therefore we can
estimate the mean energy of $\rho$ as follows:
\begin{eqnarray}\label{Tempineq}
tr(\rho H)
&=& 
\sum_j p_j E_j \leq c + 
2^n \frac{1}{e^{(c-a)/(KT)}} E_{max} \nonumber\\
&=& 
a+\frac{1}{2} (d-a) + 
\frac{e^{n\ln 2}}{e^{(d-a)/(2KT)}} E_{max}\,,
\end{eqnarray}
where $E_{max}$ is the largest eigenvalue of $H$. It is $O(n^3)$ since
the Hamiltonian contains $n(n-1)(n-2)/6$ three-body interaction terms.
The mean energy is asymptotically less than $d_n$ if $(d-a)/(2KT) > n
\ln 2$. This can be achieved by the temperature
\begin{equation}\label{Temp}
T_n< \frac{1}{K n 2\ln 2 q(n)}=:\frac{1}{K\tilde{q}(n)}\,,
\end{equation}
i.e., the temperature must only decrease as the reciprocal of a
polynomial in $n$.

If $q$ is chosen appropriately, the temperature in eq. (\ref{Temp}) is
sufficient to solve FQMA-problems provided that $n$ is the size of the
register where the meta circuit $\tilde{U}$ acts on plus the size of
clock register.

Note that a cooling mechanism is in some sense less worth than an
extremely good refrigerator. The cooling mechanism required to solve
FQMA has only to be able to bring the system in an extremely low
temperature after one has specified the Hamiltonian. It is comparable
to a fridge with some buttons to provide it with information whether
cheese, SAT or ice cream should be cooled, i.e., one specifies
\[
H_{cheese}
\hbox{
or
}
H_{SAT}
\hbox{
or
}
H_{ice\, cream} \dots
\]
A more explicit upper bound on the required temperature can easily be
derived for solving QMA-problems instead of FQMA. Assume that the
circuit either accepts all states with probability less than
$\epsilon$ or there is a state that is accepted with probability
$1-\epsilon$. Let $L$ be the length of the circuit and $H$ be the
Hamiltonian corresponding to $U$. We assume that all either there is
an eigenvalue of smaller than $\epsilon/(L+1)$ or that all eigenvalues
are larger than $(1-\epsilon)/(L+1)$. This can always be achieved by
amplification and considering instead the Hamiltonian corresponding to
the meta circuit. The number of qubits the Hamiltonian acts on is
given by $n:=n_x+m_x+L+1$ where $U$ acts on an $n_x+m_x$ dimensional
register as in Definition \ref{QMA}. We choose $T_n$ with
\begin{equation}\label{Temp2}
T_n< \frac{1-2\epsilon}{4\ln 2 K (L+1) n}\,.
\end{equation}

Set the cutoff value at
\[
c: = \frac{1+2\epsilon}{4(L+1)}\,.
\]
Then one can check using eq.~(\ref{Tempineq}) that if there is a
witness then for large $n$ the mean energy is smaller than
\[
d=\frac{1}{2(L+1)}\,.
\]
We conclude that for $T_n$ satisfying inequality (\ref{Temp2}) either the
mean energy is above $(1-\epsilon)/(L+1)$ or below $1/(2(L+1)$
depending on the existence of a witness. Since the gap between these
values is greater than the reciprocal of a polynomial they can
efficiently be distinguished using some copies of the same Gibbs
state.

Let us just mention a similar bound on the temperature which is
required to find the solution of the classical (NP-complete)
Independent Set Problem. In the Ising-Hamiltonian on $n$ spins
considered in \cite{Pawelcompass} the energy value of the ground
states encodes a graph theoretical NP-problem. The energy gap between
the ground state and the first excited state is a constant value
$\Delta E$. Then
\[
T \leq \frac{1}{\ln 2 K \Delta E n}
\]
is sufficient to have for large $n$ an almost sure answer.

This can be seen by choosing the cutoff value
\[
c:=a+\frac{1}{4}\Delta E\,.
\]
If the mean energy of the system is above
\[
a + \frac{1}{2} \Delta E
\]
we can conclude that the answer of the NP-problem is ``no'', otherwise
it is ``yes''.

\section*{Acknowledgements}
Thanks to Markus Grassl for helpful discussions. This work was
supported by the BMBF-project 01/BB01B. It was finished during P.W.'s visit
at the Institute for Quantum Information at Caltech. He appreciates the
hospitality of IQI.


\end{document}